\documentclass[twocolumn,showpacs,amssymb,preprintnumbers,nobibnotes,aps,prX,floats,psfig]
{revtex4-1}
\usepackage{textcomp,amssymb,graphicx,epsfig}
\usepackage{hyperref}
\usepackage{amsbsy}
\usepackage{amsmath,amsfonts,color}
\usepackage{IEEEtrantools}
\usepackage{float}
\usepackage[dvipsnames]{xcolor}
\date{}

\begin{document}
\title{
Charge and orbital order due to cooperative Jahn-Teller effect 
in manganite chains}

\author{Ravindra Pankaj}

\author{Sudhakar Yarlagadda}

\affiliation{
CMP Division, Saha Institute of Nuclear Physics,
Kolkata, India}

\date{\today}

\begin{abstract}
{We derive an effective Hamiltonian that takes into account the quantum nature of phonons
and models cooperative Jahn-Teller effect
in the adiabatic regime and at strong electron-phonon coupling  in  one dimension.
Our approach involves mapping a strong-coupling problem to a weak-coupling 
one by using a duality transformation. Subsequently,
a sixth-order
perturbation theory is employed in the polaronic frame of reference
where the small parameter is inversely (directly) proportional
to the coupling (adiabaticity).}
We study charge and orbital
order in  ferromagnetic manganite chains and address
the  pronounced electron-hole asymmetry in the observed phase diagram. 
In particular, at strong coupling, 
we  offer an explanation for the observed density dependence of the wavevector
of charge modulation, i.e., wavevector is proportional to (independent of)  electron
density on the electron-doped (hole-doped) side of the phase diagram of manganites.
{We also provide a picture 
for the charge and orbital order at special fillings
$ \frac{1}{2}$, $\frac{1}{3}$, $\frac{1}{4}$, and $ \frac{1}{5}$};
while focusing on the ordering controversy at  fillings $\frac{1}{3}$ and 
$\frac{1}{4}$,
we find that Wigner-crystal arrangement is preferred over bi-stripe order.

\end{abstract}


\maketitle

\section{Introduction}
Various transition-metal oxides such as manganites
\cite{Zimmermann,Larochelle,Jirak,Cheong2,Cheong3,Loudon,Milward,Radaelli1,Radaelli2,Wang,Garcia,Rodriguez,Diaz,Hervieu},
cuprates \cite{Tranquada}, nickelates
\cite{Cheong1},
cobaltates \cite{Cwik}, etc. display clear evidence for stripe-like magnetic and charge orders.
In doped Mott insulators  such
as cuprates and nickelates,
it has been argued that stripes are
generated by the competition between
the clustering tendency of the doped holes 
(in regions of suppressed antiferromagnetism)
and the long-range Coulomb interactions \cite{Emery}.
{On the other hand, properties in manganites arise as a compromise between the 
tendency of the carriers to delocalize owing to the kinetic
energy and their propensity to localize due to a strong cooperative
Jahn-Teller (CJT) effect and the antiferromagnetic
(AFM) interaction between the Mn core spins.
{\em Here, we show that
the CJT effect produces long-distance repulsion thereby
enabling stripe formation.}}

{Perovskite manganites R$_{1-x}$A$_x$MnO$_3$ (R = La, Pr, 
Sm, etc., A = Sr, Ca, 
etc.)
exhibit a zoo of exotic phases involving a variety of  spin, charge, and orbital textures/stripes.
{The stripe phases 
involve A-, C-, or CE-type
antiferromagnets \cite{tokura1}; 
they manifest charge orders that have doping dependent wavevectors above $x=0.5$ and doping 
independent wavevectors below $x=0.5$ \cite{Milward,braden}; and reveal C-type,  ferro-type, 
Wigner-crystal/bi-stripe
 orbital orders }\cite{salamon}.

Among the charge-ordered manganites, low-bandwidth manganites
such as Pr$_{1-x}$Ca$_x$MnO$_3$ (PCMO) display charge-ordering
for fairly large range of doping, namely, $0.3\leq x \leq 0.75$ \cite{
au_jirak,Zimmermann,tokura1};
whereas intermediate-bandwidth 
manganites such as La$_{1-x}$Ca$_x$MnO$_3$ (LCMO) exhibit charge-ordering for
$0.5 \leq x \leq 0.8$\cite{au_kallias,au_pissas}. At $x=0.5$, 
checker-board charge ordering is manifested in both PCMO and LCMO
with ordering wavevector $k=0.5a^*$, where $a^*$ is reciprocal lattice vector \cite{tokura1,salamon}. 
On the other hand, at $x=2/3$ and $x=3/4$
there is a controversy whether a bi-stripe order
or a Wigner-crystal order is realized by the system \cite{au_mori,Radaelli2,au_kallias,epstein}.
Furthermore, for  $0.8 \lesssim x \lesssim 0.85$ in LCMO, orbital order (without charge order)
involving $d_{z^2}$ orbitals
along ferromagnetic chains in a C-type antiferromagnet has been reported \cite{au_kallias}.
Additionally, at $0.3 < x < 0.5$ in Pr$_{1-x}$Ca$_x$MnO$_3$, it
has been claimed that
CE-type checker-board order (corresponding to
$x=0.5$) is retained with excess electrons 
occupying the Jahn-Teller compatible 
$d_{z^2}$ orbitals
at the empty sites of the checker board \cite{tokura1}.
{\em Here, we present a scenario for Wigner-crystal states at
$x = 2/3$ and $x=3/4$ and a C-AFM state for $x\gtrsim 0.8$.
Furthermore, we also offer an explanation for the Jahn-Teller compatible states 
realized for $x<  0.5$ in narrow-band compounds such as PCMO.}

{
Evidence of sizeable local Jahn-Teller distortions indicating strong electron-phonon
coupling has been provided in manganites  by
 direct techniques such as extended x-ray
absorption fine structure \cite{bianc} and
 pulsed neutron diffraction 
\cite{louca}.
For a long time, charge ordering in the overdoped regime ($x>0.5$) 
was considered arising from ordered arrangement of ${\rm Mn^{3+}}$ and ${\rm Mn^{4+}}$ ions,
i.e., stripes of localized charges \cite{goodenough};
for arbitrary dopings,
charge-ordering was thought to be a fine mixture of two adjacent commensurate configurations according to the 
lever rule \cite{au_mori}.
This scenario (purported to result from strong-coupling)
has been questioned based on
experiments which show that the
charge modulation continues to be uniform when passing from
commensurate to incommensurate filling \cite{Loudon,Milward}.
{\em In this article, 
we show that even strong coupling at an incommensurate filling
 produces a 
{finite peak} in the structure factor 
with 
wavevector  
that is linearly dependent on filling 
{$\nu~(=1-x)$ and rules out charge stacking faults.} }
}

{
{As regards 
theoretical efforts,
double exchange, superexchange, and
Coulombic repulsion  
have been 
utilized to study manganites.
Jahn-Teller effect (without cooperativity) was used additionally to explain  
colossal magnetoresistance  in the conducting regime of the 
manganites \cite{au_millis,
tvr1}.
On the other hand, a number of experiments suggest  Jahn-Teller effect, with cooperativity,
to be crucial in stabilizing charge ordering
in LCMO (at $x \ge 0.5$) and PCMO (at $x> 0.3 $)
\cite{au_zheng1,au_zheng2,au_tong,epstein,Zimmermann}. 
However, an effective Hamiltonian that models 
CJT effect 
is yet to be developed; consequently, there does not exist a  unified picture that
explains the rich phase diagram of manganites.}
{\em 
In the present 
work, employing perturbation theory up to sixth order, we obtain an effective Hamiltonian 
that models CJT effect (by taking into account  quantum  phonons)
in the intermediate- and narrow-band manganites in one dimension.}
In these lower tolerance materials,
the small parameter is sufficiently small
since it is directly 
proportional to the adiabaticity which is not too large and inversely
proportional to the electron-phonon coupling which is large
due to large polaronic distortion \cite{littlewood}.
Our study is 
relevant to the insulating regime $x \ge 0.5$ ($x \ge 0.3$)
of intermediate- (narrow)-band manganites where 
either zigzag or straight ferromagnetic
chains are antiferromagnetically coupled. 
{\em In our  one-dimensional model,  we  
{demonstrate
particle-hole
asymmetry by  obtaining the 
observed 
CDW-wavevector dependence on density}} 
\cite{Milward,braden}.
}

\section{Effective polaronic Hamiltonian }
\begin{figure}
\begin{center}
\includegraphics[scale=1,angle=0]{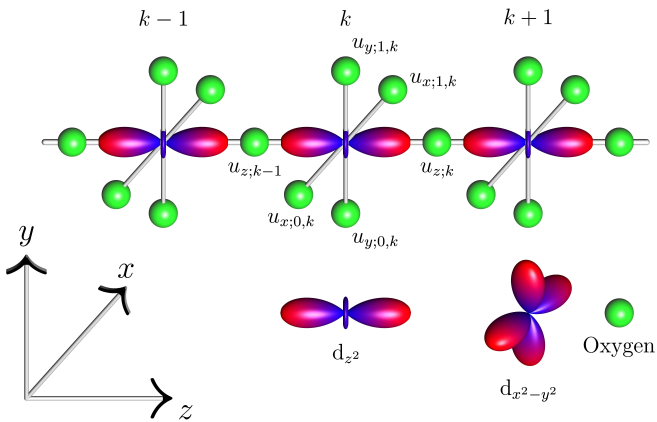}
\caption{(Color online) Depiction of 
one-dimensional cooperative  Jahn-Teller interaction in a chain
involving $d_{z^2}$ and $d_{x^2-y^2}$ orbitals and oxygen ions.
For simplicity, only  $d_{z^2}$ orbitals and oxygen ions
are displayed in the chain and their locations are indicated. 
}
\label{fig:cartoon}
\end{center}
\end{figure}

We consider a one-dimensional  Jahn-Teller chain with cooperative electron-phonon 
interaction along the z-direction and non-cooperative  electron-phonon interaction 
(of the Holstein-type \cite{au_holstein,au_sdadys})
along the x- and
y-directions as shown in Fig.~\ref{fig:cartoon}. 
We consider spinless fermions so as to model
the physics of ferromagnetic chains. We write
the Hamiltonian for cooperative-Jahn-Teller chain
as follows 
{(see Appendix \ref{app:gen_ham} for details):}
\begin{IEEEeqnarray}{rCl}
\!\!\!\!\!\!\!\! H^{CJT} =& -&t \sum_k (d^\dagger_{1,k+1}d_{1,k}+ {\rm H.c.}) \nonumber \\ 
&-&g \omega_0 \sum_k \bigg [(a^\dagger_{k}+a_{k})(n_{1,k}-n_{1,k+1}) \nonumber \\
&&\qquad ~~~~+  \frac{1}{2} (b^\dagger_k+b_k) (n_{1,k}+3 n_{2,k} ) \nonumber \\
&& \qquad ~~~~ - \frac{\sqrt{3}}{2} (c^\dagger_k+c_k) (d^\dagger_{1,k}d_{2,k}+ {\rm H.c.})\bigg ] \nonumber \\
&+& \omega_0 \sum_k (a^\dagger_{k}a_{k} + b^\dagger_k b_k + c^\dagger_k c_k), 
\label{eq_HCJT_sq}
\end{IEEEeqnarray}
where $n_{1,k} \equiv d^\dagger_{1,k}d_{1,k} $ and $n_{2,k} \equiv d^\dagger_{2,k}d_{2,k} $
with $d^\dagger_{1,k}$ $(d^\dagger_{2,k})$ being the creation operator for $d_{z^2}$ 
$(d_{x^2-y^2})$ orbital; phonon creation and annihilation operators are
defined as follows:
\begin{IEEEeqnarray*}{rCl}
\frac{a^\dagger_{k}+a_{k}}{\sqrt{2M\omega_0}} &=& u_{z;k} ,\\
 \frac{b^\dagger_k+b_k}{\sqrt{2\frac{M}{4}\omega_0}}
&=& (u_{x;1,k}-u_{x;0,k})+(u_{y;1,k}-u_{y;0,k}), \\
\frac{c^\dagger_k+c_k}{\sqrt{2\frac{M}{4}\omega_0}}
&=&(u_{x;1,k}-u_{x;0,k})-(u_{y;1,k}-u_{y;0,k}),
\end{IEEEeqnarray*}
where $u_x$, $u_y$, and $u_z$ are, respectively, displacements of the oxygens
along $x$-, $y$-, and $z$-axes (See Fig.~\ref{fig:cartoon}).

We will now 
modify the 
Lang-Firsov transformation \cite{au_lang} and apply it to the above Hamiltonian so that we can
perform perturbation in the polaronic (Lang-Firsov transformed) frame of reference. The transformed Hamiltonian is 
given by $\tilde{H}^{CJT} = \exp(S) H^{CJT} \exp(-S)$
where
\begin{IEEEeqnarray}{rCl}
S =& -&g \sum_k [(a^\dagger_{k}-a_{k})(n_{1,k}-n_{1,k+1}) \nonumber \\
&+& \frac{1}{2} (b^\dagger_k-b_k) (n_{1,k}+3 n_{2,k} )]. 
\label{op_lang}
\end{IEEEeqnarray}
Here, in our modified Lang-Firsov transformation, it should be noted that we have included only the density terms 
and ignored the orbital-flip terms  ($d^\dagger_{1,k}d_{2,k}$ and its Hermitian conjugate) appearing in the interaction
part of the above equation (\ref{eq_HCJT_sq}). This choice is dictated by mathematical expediency
to arrive at an analytic expression. 
Then, the Lang-Firsov transformed Hamiltonian is given by 
$\tilde{H}^{CJT} = H_{\rm ph} +H_{\rm s}+ H_1$ where
\begin{eqnarray}
H_{\rm ph} = \omega_0 \sum_k (a^\dagger_{k}a_{k} + b^\dagger_k b_k + c^\dagger_k c_k) ,
\end{eqnarray}
and
\begin{eqnarray}
H_{\rm s} = 
&& - t e^{-(E_p+V_p)/\omega_0} \sum_k (d^\dagger_{1,k+1}d_{1,k}+ {\rm H.c.})
\nonumber  \\
&& - E_p \sum_k (n_{1,k}  
+ n_{2,k}) 
+ 2 V_p \sum_k n_{1,k}n_{1,k+1}, 
\label{eq_H_0}
\end{eqnarray}
{with   $E_p = \frac{9}{4} g^2 \omega_0$ being the  polaronic energy
and $2V_p = 2g^2 \omega_0$ being the repulsion between nearest-neighbor (NN)
$d_1$-electrons due to cooperative interaction. In Eq. (\ref{eq_H_0}), it is important
to note that there is no interaction between  NN $d_2$-electron and $d_1$-electron or
between two NN $d_2$-electrons.}
The remaining term is the perturbation:
\begin{eqnarray}
\!\!\!\!\!\!\!\!
H_1 
\approx -t e^{-\frac{E_p+V_p}{\omega_0} }
 \sum_k [ d^\dagger_{1,k+1}d_{1,k}
\{
{\cal{T}}_{+}^{k \dagger} {\cal{T}}^{k}_{-}
-1 \} + {\rm H.c.}],
\label{eq_H_1_pert}
\end{eqnarray}
where ${\cal{T}}^{k}_{\pm} \equiv \exp[\pm g( 2 a_{k} - a_{k-1} - a_{k+1})\pm \frac{g}{2}(b_k-b_{k+1})]$.
The details of the exact transformation along with the perturbation theory are given in
 Appendix \ref{app:IIorder}. For realistic values of adiabaticity and electron-phonon coupling in manganites, 
we need to retain dominant terms up to sixth order in perturbation as will be explained
below. 
{Now, while performing perturbation theory, the NN repulsion between two $d_1$-electrons
must be carefully accounted for; hence, in the  dominant-interaction 
processes considered in Fig.~\ref{fig_sixth_pertur}, we differentiate between situations where
 the mobile $d_1$-electron does not interact interact with another $d_1$-electron
 [see Figs.~\ref{fig_sixth_pertur}(c),  \ref{fig_sixth_pertur}(e), \ref{fig_sixth_pertur}(g)]
 and those where the mobile $d_1$-electron does interact with another $d_1$ electron
 [see Figs.~\ref{fig_sixth_pertur}(d),  \ref{fig_sixth_pertur}(f), \ref{fig_sixth_pertur}(h)].}

The second-order term reads as follows:
\begin{eqnarray}
&& H_{\rm eff}^{II} = \sum_{m}
\frac{\langle 0|_{ph} H_{1} |m\rangle_{ph}
 \langle m|_{ph} H_{1} |0\rangle_{ph}}
{E_{0}^{ph} - E_{m}^{ph}} 
\nonumber \\
&& = 
- \frac{t^2}{E_p+2 V_p }e^{-\frac{E_p}{\omega_0} } 
\sum_k P_{k+1} \left [
d^\dagger_{1,k+2}
d_{1,k}
+ {\rm H.c.}\right ]
\nonumber \\
&& \quad -  \frac{t^2}{2E_p+2V_p}
\sum_k P_{k+1}
\big[ n_{1,k} (1- n_{1,k+2}) 
\nonumber \\
&& ~~~~~~~~~~~~~~~~~~~~~~~~~~~~~~~~~ + n_{1,k+2} (1- n_{1,k})
\big]
\nonumber \\
&& \quad -  \frac{t^2}{2E_p+4V_p}\sum_k P_{k+1}
\big[  n_{1,k}   n_{1,k+2} + n_{1,k+2}  n_{1,k}
 \big] ,
\label{eq_H^2_1_2}
\end{eqnarray}
where $P_{k+1} \equiv (1-n_{1,k+1}) (1-n_{2,k+1})$ projects out electrons at site $k+1$.
{In the above equation, on the right-hand side (RHS),
the 
expression containing $d^\dagger_{1,k+2}d_{1,k}$ in
the first term 
represents next-nearest-neighbor (NNN) hopping as displayed in 
Fig.~\ref{fig_sixth_pertur}(a)
(see Appendix \ref{app:IIorder} and Ref. \onlinecite{sahinur1} for details).
It should be noted that Fig.~\ref{fig_sixth_pertur}(b) does not contribute to 
{the second-order term since NN}
repulsion is large.
The 
expression containing $n_{1,k} (1- n_{1,k+2})$ in the
second term on the RHS corresponds
to  Fig.~\ref{fig_sixth_pertur}(c) and Figs.~\ref{fig_pertur_process}(a) and \ref{fig_pertur_process}(b);
here, NNN site is unoccupied by $d_1$-electron and the lattice distortion
remains unchanged while the electron hops to the neighboring site and returns back.
The  denominator $2E_p+2V_p$ in the
second term on the RHS  is the difference of the energies
of the intermediate state and the initial state; the origin of the denominator is explained as follows.
The initial state shown in Fig.~\ref{fig_pertur_process}(a)
has energy  $-E_p$ whereas
the intermediate state depicted in Fig.~\ref{fig_pertur_process}(b) has 
 energy $E_p+2V_p$; 
in the energy of the intermediate state,  $+E_p$ arises due to the distortion without
the electron whereas $2V_p$ contribution is from the repulsion between the electron and the
oxygen ion displaced towards it.
Next, 
the 
expression containing $n_{1,k} n_{1,k+2}$ in the
the last term on the RHS is depicted in
Fig.~\ref{fig_sixth_pertur}(d) and Figs.~\ref{fig_pertur_process}(c) and \ref{fig_pertur_process}(d)
with NNN site being occupied by $d_1$-electron; here, the
energy of the intermediate state [Fig.~\ref{fig_pertur_process}(d)] is $2E_p+4V_p$ 
above the ground state with $4V_p$ representing
repulsion felt by the electron at site $k+1$ due to  neighboring  oxygens displaced towards it
on both the sides.
The last 
two terms on the RHS indicate repulsion between NNN electrons only if
no electron is present between them.
Here it should be noted that, while carrying out perturbation theory,
we assumed $t e^{-(E_p+V_p)/\omega_0} << \omega_0$ which is valid for manganites.
Furthermore, the small parameter
of our perturbation theory is $\sqrt{\frac{t^2}{(2E_p+2V_p)\omega_0}} = \sqrt{\frac{2}{13}}\frac{t}{g \omega_0  }$;
it is obtained from the following
largest coefficients in $2l$-order processes which involve 
the electron 
hopping $l$ times back and forth between NN sites
while NNN site is not occupied by $d_1$-electron [for a similar analysis for a Holstein model,
 see Ref. \onlinecite{au_rpys}]:
\begin{eqnarray*}
\Bigg(\frac{t^2}{(2E_p+2V_p)\omega_0} \Bigg)^l \omega_0
\sum_k P_{k+1}
\big[ &&  n_{1,k} (1- n_{1,k+2}) 
\nonumber \\
&& 
+ n_{1,k+2} (1- n_{1,k})
\big] .
\end{eqnarray*}
Thus we 
have shown that {\em the polaronic (Lang-Firsov) transformation is actually a duality transformation
that maps the original strong-coupling problem in
Eq.~\eqref{eq_HCJT_sq} {\rm[}with perturbation proportional
to $(g\omega_0)/t${\rm]} 
to a weak-coupling problem {\rm[}with small
parameter proportional to $t/(g \omega_0)${\rm]}}
\cite{sanjoy_ys}.

The dominant contribution for the next-to-next-nearest neighbor (NNNN) interaction is given by 
fourth-order processes 
and expressed below 
{(for clarity on the associated lattice distortions,
see Fig. \ref{fig_fourthorder} in Appendix \ref{IV-process})}:
\begin{eqnarray}
&& \!\!\!\!\!\!\!\!\!\!  H_{\rm eff}^{IV}= 
- \frac{t^4}{ \bigl(2E_p+2V_p \bigr)^2 
\bigl(2 E_p\bigr) }
\nonumber \\
&& ~~~~~ \times
\sum_k P_{k+1}P_{k+2}\biggl[ n_{1,k}\Bigl(1- \frac{V_p}{E_p+V_p} n_{1,k+3} \Bigr)
\nonumber \\
&& ~~~~~~~~~~~~~~~~~~~~~~~~~~+n_{1,k+3}\Bigl(1- \frac{V_p}{E_p+V_p} n_{1,k}\Bigr)\biggr].
\end{eqnarray}
In the above equation, the expression containing 
$ n_{1,k}\Big(1- \frac{V_p}{E_p+V_p} n_{1,k+3} \Big) $ is obtained
by considering the processes depicted in Figs.~\ref{fig_sixth_pertur}(e) and \ref{fig_sixth_pertur}(f).
We first note that, if an oxygen is displaced  towards the new location of the mobile electron in any intermediate state,
the energy of the intermediate state is enhanced further by $2V_p$  with respect to the ground state
[see Figs. \ref{fig_fourthorder}(b), \ref{fig_fourthorder}(e), and  \ref{fig_fourthorder}(f)].
Regarding the virtual hopping to the site $k+1$ in Fig.~\ref{fig_sixth_pertur}(e)
or in Fig.~\ref{fig_sixth_pertur}(f), the intermediate state has energy
$2E_p + 2V_p$ above the ground state.
Next, associated with virtual
hopping to the site $k+2$ in Fig.~\ref{fig_sixth_pertur}(f) [Fig.~\ref{fig_sixth_pertur}(e)], 
the intermediate state has energy $2E_p+2V_p$ [$2E_p$] above the ground state.

Lastly, the sixth-order processes leading to the dominant contribution for the 
next-to-next-to-next-nearest neighbor (NNNNN) repulsion are depicted in 
Figs.~\ref{fig_sixth_pertur}(g) and \ref{fig_sixth_pertur}(h) and yield the following expression:
\begin{eqnarray}
&& \!\!\!\!\!\!\!\!\!\!\!\!\!  H_{\rm eff}^{VI}= 
 - \frac{t^6}
{ \bigl(2E_p+2V_p \bigr)^2 \bigl(2 E_p\bigr)^3 }
\nonumber \\
&& ~~~ \times
\sum_k \prod_{i=1,2,3}P_{k+i}
\biggl[ n_{1,k}\Bigl(1- \frac{V_p}{E_p+V_p} n_{1,k+4} \Bigr)
\nonumber \\
&&~~~~~~~~~~~~~~~~~~~~~~~~~~~+n_{1,k+4}\Bigl(1- \frac{V_p}{E_p+V_p} n_{1,k}\Bigr)\biggr].
\label{eq_eff_hamVI}
\end{eqnarray}
Then, up to sixth order in perturbation, the effective Hamiltonian 
for the CJT chain is given by
\begin{eqnarray}
 H^{CJT}_{\rm eff} = H_{\rm s} +H^{II}_{\rm eff} + H^{IV}_{\rm eff} + H^{VI}_{\rm eff} .
 \label{eq_eff_ham}
\end{eqnarray}

{Interestingly, in our effective Hamiltonian,
presence of an in-between electron completely blocks the repulsion
between  the two 
surrounding electrons, i.e., screening is 100\% in contrast to long-range Coulomb 
repulsion.}
{Here, it should also be pointed out that odd order (such as third or higher order)
in perturbation theory leads to hopping terms that are negligible compared to
all the terms (including the NN and NNN hopping terms)
in Eq. (\ref{eq_eff_ham}).}

\begin{figure}
\includegraphics[scale=1,angle=0]{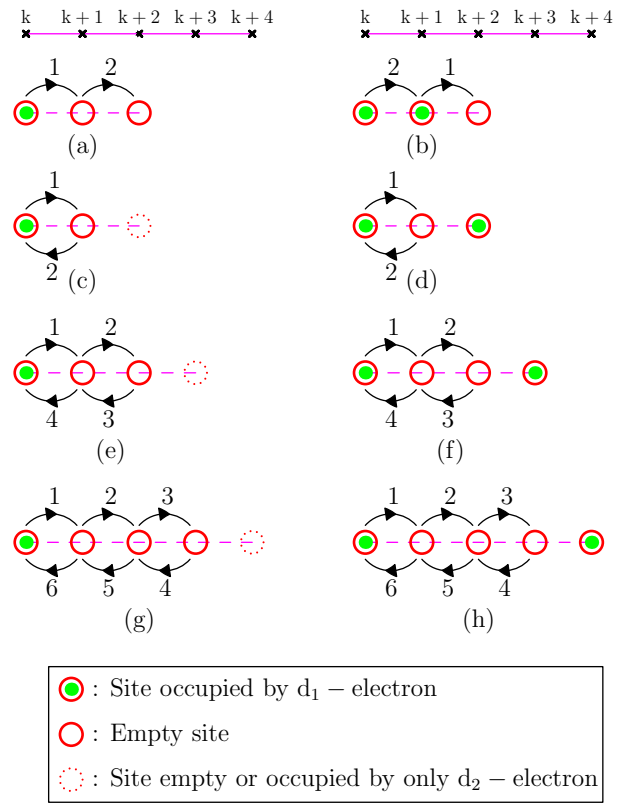} 
\caption{(Color online) Schematic depiction of processes yielding dominant interaction terms in perturbation theory.
In second-order perturbation, (a) a single particle hopping forward twice; (b) two particles
sequentially hopping forward;
(c) [(d)] a particle hopping 
to NN site and returning while NNN site is unoccupied [occupied] by $d_1$-electron.
In fourth-order perturbation, 
(e) [(f)] a particle hopping 
to NNN site and coming back while NNNN site is unoccupied [occupied] by $d_1$-electron.
In sixth-order perturbation, 
(g) [(h)] a particle hopping 
to NNNN site and coming back while NNNNN site is unoccupied [occupied] by $d_1$-electron.
The numbered arrows indicate the order of hopping.}
\label{fig_sixth_pertur}
\end{figure}

\begin{figure}
\includegraphics[scale=1,angle=0]{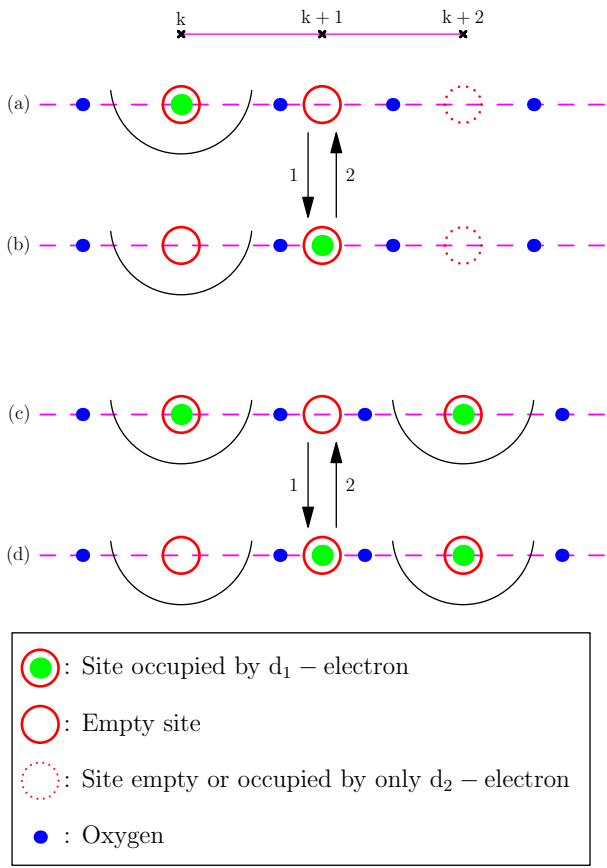} 
\caption{(Color online) Display of initial/final and intermediate states with concomitant
lattice distortions in a second order perturbation
process with electron hopping 
 to NN site and coming back. 
When NNN site is unoccupied by $d_1$-electron: 
(a) initial/final state and (b) intermediate state.
When NNN site is occupied by $d_1$-electron: 
(c) initial/final state and (d) intermediate state.
The order of hopping is specified by the numbered arrows.
}
\label{fig_pertur_process}
\end{figure}

\section{Analysis of CJT model}

Owing to the large on-site inter-orbital repulsion, 
there exists only three
possibilities, i.e., site is unoccupied or occupied by either a $d_1$-electron or a $d_2$-electron.
The size of the Hilbert space is $3^N$ where $N$ is the total number of sites. 
However, for a fixed number $N_1$ of $d_1$-electrons  and $N_2$ of $d_2$-electrons,
it further reduces to 
$^N C_{N_p} \times ^{N_p}C_{N_1}$ with $N_p=N_1+N_2$ being the 
total number of particles.
{We diagonalize the effective Hamiltonian in Eq.~\eqref{eq_eff_ham} using modified Lanczos algorithm \cite{au_gagliano} 
for fixed values of $N$, $N_p $, and $N_1$  to obtain the minimum energy state of the system 
when electron-phonon coupling
$g=2.3$ and adiabaticity $\frac{t}{\omega_0} =3.0$.}  

To identify the charge and orbital ordering, we study the correlations of the particles. 
The two-point correlation function for density fluctuations of $d_{a (b)}$-electrons
with $a(b)=1,2$
is given by
\begin{IEEEeqnarray*}{rCl}
 W_{d_{a}d_{b}}(l)
 =\frac{4}{N}\sum_j \left[\langle n_{a,j} n_{b,j+l} \rangle-\langle n_{a,j}\rangle \langle n_{b,j+l}\rangle\right] ,
\label{eq_Wuu}
\IEEEyesnumber
\IEEEeqnarraynumspace
\end{IEEEeqnarray*}
where 
$\langle n_{a,j}\rangle =\frac{N_a}{N}$.
  Then, the observable structure factor is expressed as the Fourier transform of $W_{d_{a}d_{b}}(l)$ as follows:
\begin{IEEEeqnarray*}{rCl}
 S_{d_{a}d_{b}}(k)=\sum_l e^{ikl} W_{d_{a}d_{b}}(l) ,
\label{eq_Suu}
\IEEEyesnumber
\IEEEeqnarraynumspace
\end{IEEEeqnarray*}
where the wavevector $k=\frac{2 n^\prime \pi}{N}$ with 
{$n^\prime=1,2,\ldots,N$ and lattice constant taken to be
of unit length}.

\subsection{Up to half-filling case}

\begin{figure}
\includegraphics[scale=1,angle=0]{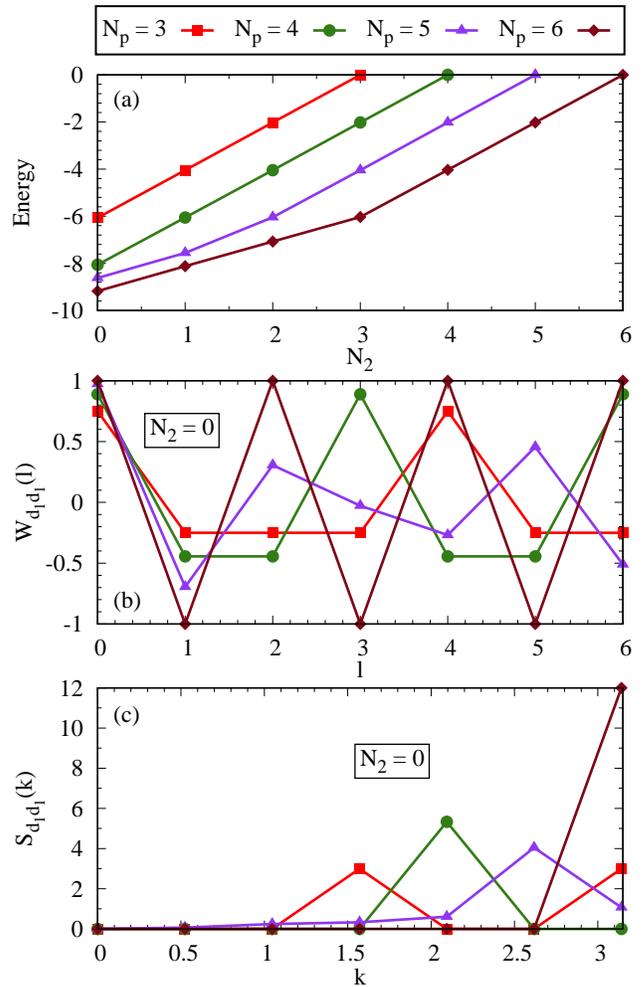}
\caption{(Color online)
 Plots  of (a) lowest energy (in units of $\frac{t^2}{2E_p+2V_p} \Big )$
 excluding the polaronic energy;
 (b) correlation function ${ W_{d_{1}d_{1}}(l)}$; 
 and  (c) structure factor ${  S_{d_{1}d_{1}}(k)}$
at filling  $\nu \le 1/2$
in a straight chain with periodic boundary condition.
 The chain has $N=12$ sites and ${ N_p}$ electrons 
 {of which ${ N_2}$ are ${ d_2}$-electrons};
electron-phonon coupling $g=2.3$ while  adiabaticity $\frac{t}{\omega_0}=3.0$.
At $\nu= 1/n$, $W_{d_{1}d_{1}}(l=m/\nu)=4\nu(1-\nu)$
 and $W_{d_{1}d_{1}}(l\neq m/\nu)=-4\nu^2$; and
we get charge order with
$S_{d_{1}d_{1}}(k )= 4 N_p^2 /N$ at the reported
wavevector $k =2 \pi \nu$ \cite{Milward,braden} and 0 otherwise.
}
\label{fig_energy_12_tw3}
\end{figure}

We display lowest energies
of the CJT system as a function of 
number of $d_2$-electrons in Fig.~\ref{fig_energy_12_tw3}(a). 
We observe from this figure that, for the ground state up to half filling,
electrons occupy only the $d_{z^2}$ orbitals whereas $d_{x^2-y^2}$ orbitals
remain unoccupied. Hence, up to half filling, i.e., $\nu=\frac{N_p}{N} \leq \frac{1}{2}$,
we need to consider only one-orbital case for further study.
{Consequently, in the ground state, $N_p= N_1$ as $N_2 = 0$ and the Hilbert 
space reduces to $^NC_{N_1}$ for a fixed number of particles. Instead of using modified Lanczos
algorithm,
we do full exact diagonalization  
of the effective Hamiltonian in Eq.~\eqref{eq_eff_ham} so as to avoid problems due to degeneracy; 
we calculate correlation functions, structure factor, and excitation energies.}

{At strong coupling and in the adiabatic regime which are relevant to the 
manganites, we will analyze correlation function and structure factor for the ground state.} 
We display the two-point correlation function in Fig.~\ref{fig_energy_12_tw3}(b);
 we
 observe that the system
 at strong coupling and in 
the adiabatic regime has oscillatory correlation function with fixed 
amplitude for special fillings such as $\frac{1}{2}$, $\frac{1}{3}$, and
$\frac{1}{4}$.
At the above mentioned special fillings, from the excitation gaps in Table. I and Fig.~\ref{fig_energy_12_tw3}(b),
it is evident that the
 system has an insulating CDW state 
with ordering wavevector $2 \pi \nu$ as depicted by structure factor peaks
in Fig.~\ref{fig_energy_12_tw3}(c).
For special fillings, 
{as will be explained below,
$S_{d_{1}d_{1}}(k)$ peaks attain their maximum value for $k=2 \pi \nu$ and zero
everywhere else when
the particles 
are strongly localized
at their respective periodic positions in the lattice.}

Away from the above mentioned special fillings, system 
{does not have long range order}
as reflected by the 
{not so regular} behavior of the correlation function
in Fig.~\ref{fig_energy_12_tw3}(b).
Away from special fillings, system has short range correlations corresponding to the 
wavevector $2 \pi \nu$ as shown by the
peak in the structure factor in Fig.~\ref{fig_energy_12_tw3}(c).
In contrast to special fillings, structure factor for non-special fillings is not zero away from $k = 2 \pi \nu$.
{
{As will be shown below, at $\nu \neq 1/n$ with $n$ being an integer,
the system is metallic in the absence
of disorder; whereas, in the presence of even weak
disorder, the system becomes insulating with a 
{finite peak expected } in the structure factor
at $k = 2 \pi \nu$}}.

At special fillings $\nu =1/n$ with $n$ being an integer,
 when system is in a charge ordered state  
with particles separated by distances $m/\nu$,
it is interesting to note that we can  
simplify Eqs.\eqref{eq_Wuu} and \eqref{eq_Suu} in the  following  way.
Since particles are absent at a distance $l \neq \frac{m}{\nu}$ 
where $m=1, \ldots, N\nu$, 
$\langle n_{1,j} n_{1,j+l} \rangle = 0$.
Therefore, Eq. \eqref{eq_Wuu} reduces to
\begin{IEEEeqnarray*}{rCl}
W_{d_{1}d_{1}}\Big(l \neq \frac{m}{\nu} \Big) = - \frac{4 N_p^2}{N^2} = -4 \nu^2.
\label{eq_Wuu_empty}
\IEEEyesnumber
\IEEEeqnarraynumspace
\end{IEEEeqnarray*}
When particles are present at a distance $l= \frac{m}{\nu}$, 
$\sum_j \langle n_{1,j} n_{1,j+l} \rangle = N_p$. 
Hence, Eq. \eqref{eq_Wuu} simplifies to
\begin{IEEEeqnarray*}{rCl}
W_{d_{1}d_{1}}\Big(l= \frac{m}{\nu} \Big) = 4 \nu (1-\nu).
\label{eq_Wuu_filled}
\IEEEyesnumber
\IEEEeqnarraynumspace
\end{IEEEeqnarray*}
Now, the contribution of the correlation term $\sum_j \langle n_{1,j} n_{1,j+l} \rangle$
to the structure factor is given by [see Eqs.~(\ref{eq_Wuu}) and (\ref{eq_Suu})]
\begin{eqnarray}
S^c_{d_1d_1}(k)&=&\frac{4}{N}\sum_l e^{ikl}\sum_j \langle n_{1,j} n_{1,j+l} \rangle 
\nonumber \\
  &=&   \frac{4}{N}\sum_{m=1}^{N\nu} e^{ik\frac{m}{\nu}} N \nu . 
\label{Sc}
\end{eqnarray}
On the other hand, contribution of the mean-field term $\sum_j \langle n_{1,j}\rangle \langle n_{1,j+l} \rangle$
to the structure factor is 
\begin{eqnarray}
\!\!\!\!\!\!\!\!\!\!
S^m_{d_1d_1}(k)&=&\frac{4}{N}\sum_l e^{ikl}\sum_j \langle n_{1,j}\rangle \langle n_{1,j+l} \rangle
\nonumber \\
&=&  4 N \nu^2  \delta_{k,0} .
\label{Sm}
\end{eqnarray}
Next, from Eq.~(\ref{Sc}), we note that $S^c_{d_1d_1}(k=2\pi\nu p) =4N\nu^2$ and $S^c_{d_1d_1}(k\neq 2\pi\nu p) =0$ 
for integer values of $p$ with $0 \le p < n$.
Since $S_{d_1 d_1}(k)=S^c_{d_1d_1}(k)-S^m_{d_1d_1}(k)$, at $\nu =1/n$, it  follows that 
\begin{IEEEeqnarray*}{rCl}
 S_{d_{1}d_{1}}(k =2 \pi \nu p) = 4 N \nu^2 (1-\delta_{k,0}) ,
\label{eq_Suu_peak}
\IEEEyesnumber
\IEEEeqnarraynumspace
\end{IEEEeqnarray*}
where $4 N \nu^2$ is the maximum possible value for $S_{d_{1}d_{1}}(k)$; it also follows that
\begin{IEEEeqnarray*}{rCl}
 S_{d_{1}d_{1}}(k \neq 2 \pi \nu p)= 0 .
\label{eq_str_not2pinu}
\IEEEyesnumber
\IEEEeqnarraynumspace
\end{IEEEeqnarray*}

At special fillings
$\nu = \frac{1}{2}, \frac{1}{3}$, and $\frac{1}{4}$,
correlation function values,  obtained in Eqs.~\eqref{eq_Wuu_empty} and \eqref{eq_Wuu_filled}
for CDW state, are in complete agreement with those 
in Fig.~\ref{fig_energy_12_tw3}(b);
furthermore, Fig.~\ref{fig_energy_12_tw3}(c) 
exactly matches with 
Eq.~\eqref{eq_Suu_peak} at wavevector $k=2\pi\nu $ and 
with Eq.~\eqref{eq_str_not2pinu} when $k \neq 2\pi\nu $.
{The fact that the structure factor peaks at $k=2\pi\nu $ at all fillings
(including $\nu \neq 1/n$)
is in agreement
with experimental observations \cite{Milward,braden}. Furthermore, at $\nu = 1/n$,
the peaks at $k=2\pi\nu $ attain their maximum possible value
(indicating  charge order)
similar to the case
 in Fig.~4 of Ref. \onlinecite{Loudon} where the peak at 
$k=2\pi\nu $ is at its allowed  maximum by being approximately equal to the peak at $k=0$.
Next, 
at $\nu \neq 1/n$,
 peak values
are sizeably smaller than the maximum possible value 
$4N\nu^2$ much like 
the situation in Ref. \onlinecite{Loudon}
where peak values at $k=2\pi\nu $ are much smaller
than those at $k=0$. Here, it should be pointed out that our structure factor at $k=0$
becomes zero and not its maximum  [as in Fig.~4 of Ref. \onlinecite{Loudon}] 
because in the definition of
$S(k)$ [in Eq.(\ref{eq_Suu})] we subtracted the mean-field term 
$S^m_{d_1d_1}(k)=\frac{4}{N} \sum_{j,l} e^{ikl} \langle n_{1,j}\rangle \langle n_{1,j+l}\rangle=4N\nu^2 \delta_{k,0}$.}

At the special filling $\nu =1/2$, NNN repulsion, NNNN repulsion, and NNNNN repulsion, 
in the total effective
Hamiltonian of Eq.~\eqref{eq_eff_ham}
are small compared  to the strong NN repulsion term.
Therefore, it is the NN and NNN hopping terms  in  Eq.~\eqref{eq_eff_ham} that
compete with the NN repulsion term. For realistic values of adiabaticity and coupling 
relevant to the manganites, NN repulsion is extremely large compared to either of the
hopping terms 
{and this leads} to a CDW state at half filling. Similarly, at $\frac{1}{3}$ filling 
repulsion terms up to NNN and at $\frac{1}{4}$ filling repulsion terms up to NNNN 
will be relevant and will compete
with 
the NN and NNN hopping terms.

\begin{figure}[t]
\includegraphics[scale=1,angle=0]{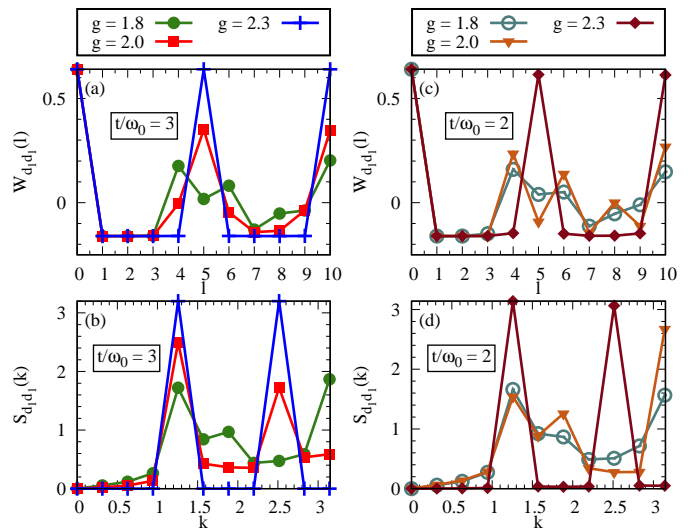}
\caption{(Color online)
{At
various values of the electron-phonon coupling $g$ and adiabaticity $t/\omega_0 $,
comparison of two-point correlation function ${ W_{d_{1}d_{1}}(l)}$ and structure factor $S_{d_{1}d_{1}}(k)$
at filling $\nu=\frac{1}{5}$ 
and $N=20$ sites
in a 
straight chain with periodic boundary condition.}
CDW occurs only at large values of $g$ with modulation wavevector $k=2\pi \nu$.
}
\label{fig_N20}
\end{figure}

We will now present arguments to show that, in the absence of disorder,
{the system is metallic away from special fillings $\nu =1/n$}.
At fillings $1/3 < \nu < 1/2$, our effective Hamiltonian in Eq.~\eqref{eq_eff_ham}
can be represented by a $t_1-t_2-V_1-V_2$
 model of spinless fermions
where $t_1$ and $t_2$ are the NN and NNN hopping terms, respectively;
$V_1$ and $V_2$  are the NN and NNN repulsion terms with $V_1 >> V_2 >> t_2 \gtrsim t_1$.
By using arguments
employed in Refs. \onlinecite{au_rpys,sahinur2},
this $t_1-t_2-V_1-V_2$
 model can be further simplified.
Due to the large NN repulsion $V_1$, with each
particle 
we associate a vacant site adjacent to it (say, 
on the right side of the particle). Then by deleting
all the   vacant sites that are adjacent on the right-side of the particles and having only a NN
repulsion $V = V_2$  in the reduced
system of $N-N_p$ sites, we get the same eigenenergy spectrum.
The effective model for Eq.~\eqref{eq_eff_ham} is the following reduced  $t_1-t_2-V$ model
\begin{eqnarray*}
&& t_1\sum_k 
( d^\dagger_{1,k+1}
d_{1,k}
+ {\rm H.c.} ) \nonumber \\
&& + t_2\sum_k P_{k+1} \! \left [
d^\dagger_{1,k+2}
d_{1,k}
+ {\rm H.c.}\right ]
+ V \sum_k   n_{1,k}   n_{1,k+1}  ,
\end{eqnarray*}
at fillings $1/2 < \nu = N_p/(N- N_p) < 1$
and with a new nearest-neighbor repulsion $V=V_2$. 
Next, at fillings $1/4 < \nu < 1/3$,
Eq.~\eqref{eq_eff_ham}
can be represented by a $t_1-t_2-V_1-V_2-V_3$
 model 
 with $V_3$ being the NNNN repulsion and $ V_2>> V_3 >> t_2 \gtrsim t_1 $;
 this model can again be reduced to the above
 $t_1-t_2-V$ model at fillings $1/2 < \nu = N_p/(N-2 N_p) < 1$ 
 and with a new nearest-neighbor repulsion $V=V_3$.
Similarly, the case of fillings $1/5 < \nu < 1/4$
can be represented by a $t_1-t_2-V_1-V_2-V_3-V_4$ model where
$V_4$   is the NNNNN repulsion with $ V_3>> V_4 >> t_2 \gtrsim t_1 $; 
here too the effective model
is the reduced $t_1-t_2-V$ model at filling $1/2 < \nu = N_p/(N-3 N_p) < 1$
and with a new nearest-neighbor repulsion $V=V_4$.
We observe that the reduced effective $t_1-t_2-V$ model
at $1/2 < \nu  < 1$ is effectively a $t_1-V$ model as NNN hopping
is not possible at $1/2 < \nu  < 1$; thus our system is a Luttinger
liquid and hence is metallic.
However, 
in the presence of even weak disorder, the effective Hamiltonian in Eq.~\eqref{eq_eff_ham}
yields an insulating behavior
 due to one-dimensionality. Here it should be
 pointed out that a source of disorder in  manganites is alkaline-earth doping. 
Lastly, to calculate
 the correlation functions, we note that  one needs to use the 
 effective Hamiltonian in Eq.~\eqref{eq_eff_ham}
 and not the reduced $t_1-t_2-V$ model.

{We will now discuss 
the effect of adiabaticity and electron-phonon coupling on CDW.
When the electron-phonon coupling $g$ and adiabaticity $t/\omega_0$
are varied 
{in the physically reasonable ranges} $1.8 \le g \le 2.3$ and $2.0 \le t/\omega_0 \le 3.0$,
the system remains in a CDW state at the special fillings $\frac{1}{2}$, $\frac{1}{3}$, 
and $\frac{1}{4}$. On the other hand, at $\nu=1/5$ and in the adiabaticity region $2.0 \le t/\omega_0 \le 3.0$, 
the system develops a CDW
only at large values of $g$ (i.e., $g \approx 2.3$) as shown in Fig. \ref{fig_N20} and Table.~I}.
{
Lastly, at fillings $\nu < 1/5$, the system will be metallic; however, introducing disorder will
make the one-dimensional system insulating.}

\begin{center}
\begin{table}[t]
{
\tiny
\begin{tabular}{|cc|cc|cc|cc|ccc|}
\hline
\multicolumn{2}{|c}{$\nu = \frac{1}{2}$} & \multicolumn{2}{|c}{$\nu = \frac{1}{3}$} & 
\multicolumn{2}{|c}{$\nu = \frac{1}{4}$} & \multicolumn{2}{|c}{$\nu = \frac{1}{5}$} & 
\multicolumn{3}{|c|}{
$\nu \neq 1/n$  gap} 

\\ [1ex] \hline 

$N$ & $\Delta_{\rm gap 
}$ & $N$ & $\Delta_{\rm 
gap}$ & $N$ & $\Delta_{\rm gap
}$ & $N$ & $\Delta_{\rm gap
}$ 
& $N$ & $N_p$ & $E_1-E_0$ \\ \hline 

 8  & 1.05882359 &
  12  & 0.47879386   &
 8  & 0.00687992  &
 10 & 0.00008676  &
 10 & 1 & 0.00000157  \\ 

 10  & 1.05882359 &
  18 & 0.47879391 &
12  & 0.00688947  &
20 & 0.00009598  &
20 & 1 & 
0.00000157
\\ 

 12  & 1.05882359 &
  &&
 16  & 0.00689015  &
  &&
  20 & 2 & 0.00000205  \\ 

 14 & 1.05882359 &
  &&
 20 & 0.00689020 &
    &&
20 & 3 & 0.00000159
\\

 16 & 1.05882359 &
   &&
   &&
   &&
   &&\\
 
 \hline 

\end{tabular}
\caption{
{Excitation gap $\Delta_{\rm gap 
}$ 
at special fillings $\nu =1/2$, $1/3$, $1/4$, and $1/5$ and lowest excitation energy ($E_1-E_0$)
at $\nu \neq 1/n$
when electron-phonon coupling $g=2.3$,  adiabaticity $\frac{t}{\omega_0}=3.0$,
and the system has $N_p$ particles in $N$ sites. 
{Here, energies are in units 
of $\frac{t^2}{2E_p+2V_p}  $.}
$\Delta_{\rm gap } >> (E_1-E_0)$ indicates CDW.  
}}}
\label{table1}
\end{table}
\end{center}

\begin{figure}[b]
\includegraphics[scale=1.0,angle=0]{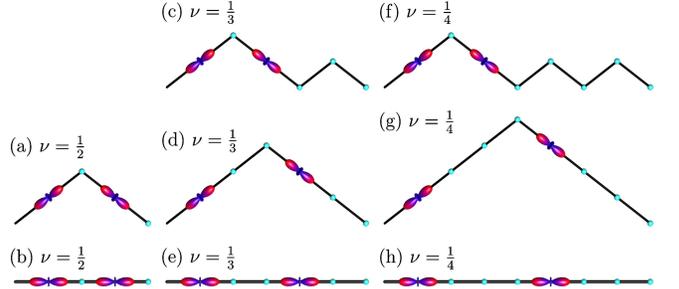}
\caption{(Color online) 
{Building blocks with 
charge/orbital } order possibilities for
intermediate- and narrow-band manganites.
Chain in (a) refers to CE-type order.
Straight chains in (b), (e), and (h) represent C-chains.
Zigzag chains in (c) and (f) correspond to bi-stripe order whereas
bent chains in (d) and (g) represent Wigner-crystal order. 
In (a), (c), (f) dominant interaction is $-\frac{t^2}{2E_p+2V_p+ \alpha_v V_p}$ with $0 < \alpha_v < 2 $; whereas,
in (d) and (g) dominant term is $-\frac{t^2}{2E_p+2V_p}$. 
Clearly, (d) is preferred over (c) while (g) is preferred over (f).
In (b) dominant interaction is $-\frac{t^2}{2E_p+2V_p+2V_p}$ implying bent chain in (a) is preferred over
straight chain in (b); similar reasoning explains why (d) [(g)] is preferred over (e) [(h)].}
\label{ferro_chains}
\end{figure}
\subsubsection{
{Types of chains and ordering at special fillings $\nu=1/n$}}
In this section, at special fillings $\nu = 1/n$,
we will compare the various possibilities depicted in Fig.~\ref{ferro_chains} for the 
charge/orbital ordered ferromagnetic chains 
that are antiferromagnetically coupled \cite{takada}.
In Figs.~\ref{ferro_chains}(a), \ref{ferro_chains}(c),and  \ref{ferro_chains}(f),
the dominant interaction is $-\frac{t^2}{2E_p+2V_p+ \alpha_v V_p}$
where $0 < \alpha_v < 2$; in fact, it can be shown that $\alpha_v = 25/32$. On the other hand,
in Figs.~\ref{ferro_chains}(d) and \ref{ferro_chains}(g), the dominant term is $-\frac{t^2}{2E_p+2V_p}$. 
Hence, clearly the chain in Fig.~\ref{ferro_chains}(d) 
 [Fig.~\ref{ferro_chains}(g)]
is energetically favorable compared
to the
chain in Fig.~\ref{ferro_chains}(c) [Fig.~\ref{ferro_chains} (f)].
Thus we see that the Wigner-crystal arrangement associated with Fig.~\ref{ferro_chains}(d)  [Fig.~\ref{ferro_chains}(g)]
should be preferred  over the bi-stripe arrangement associated with Fig.~\ref{ferro_chains}(c) 
 [Fig.~\ref{ferro_chains}(f)] at filling $\nu=1/3$ [$\nu =1/4$].

 In Fig.~\ref{ferro_chains}(b), the dominant interaction is 
$-\frac{t^2}{2E_p+2V_p+2V_p}$; hence, the bent chain in Fig.~\ref{ferro_chains}(a)
has lower energy compared to the 
straight chain in Fig.~\ref{ferro_chains}(b).
Next,  in Fig.~\ref{ferro_chains}(d)  [Fig.~\ref{ferro_chains}(g)],
the NNNN [NNNNN] interaction obtained 
due to the left-side electron virtually hopping 
(to the right and returning)
in a  fourth-order [sixth-order] perturbation
theory 
in the case of the bent chain
is given by 
$- \frac{t^4}{(2E_p+2V_p)^2(2 E_p+\alpha_v V_p)}$ 
$\Big [- \frac{t^6}{(2E_p+2V_p)^2(2E_p)^2(2E_p+\alpha_v V_p)} \Big ]$;
whereas, for the straight chain in Fig.~\ref{ferro_chains}(e) [Fig.~\ref{ferro_chains}(h)],
the corresponding NNNN [NNNNN] interaction is given by
$-\frac{t^4}{(2E_p+2V_p)^2(2 E_p+2 V_p)}$
$\Big [- \frac{t^6}{(2E_p+2V_p)^2(2E_p)^2(2E_p+2V_p)} \Big]$.
Thus, we see that the Wigner-crystal order 
pertaining to Fig.~\ref{ferro_chains}(d) [Fig.~\ref{ferro_chains}(g)]
is energetically preferred over the C-AFM state in Fig.~\ref{ferro_chains}(e) [Fig.~\ref{ferro_chains}(h)].

Lastly,
at $\nu = 1/5$, if the NNN hopping  term is  more  important than 
the  interaction from sixth-order perturbation $- \frac{t^6}{(2E_p+2V_p)^2(2E_p)^3}$,
the  energy (essentially from the kinetic part)
is lower for the straight chain than for the bent chain because
the NNN hopping over the bend is  smaller (by a factor of half) compared to the NNN hopping along the 
straight chain.  
{Consequently, at $\nu = 1/5$, the straight chain 
(without CDW order) is preferred over the bent chain;
additionally,  straight chain without CDW will continue to
be preferred even for $\nu < 1/5$ 
and we get C-AFM order (instead of a Wigner crystal) at $\nu \le 1/5$ as witnessed in LCMO \cite{au_kallias}.
}

\subsection{Above half-filing case}

\begin{figure}[b]
\includegraphics[scale=0.9,angle=0]{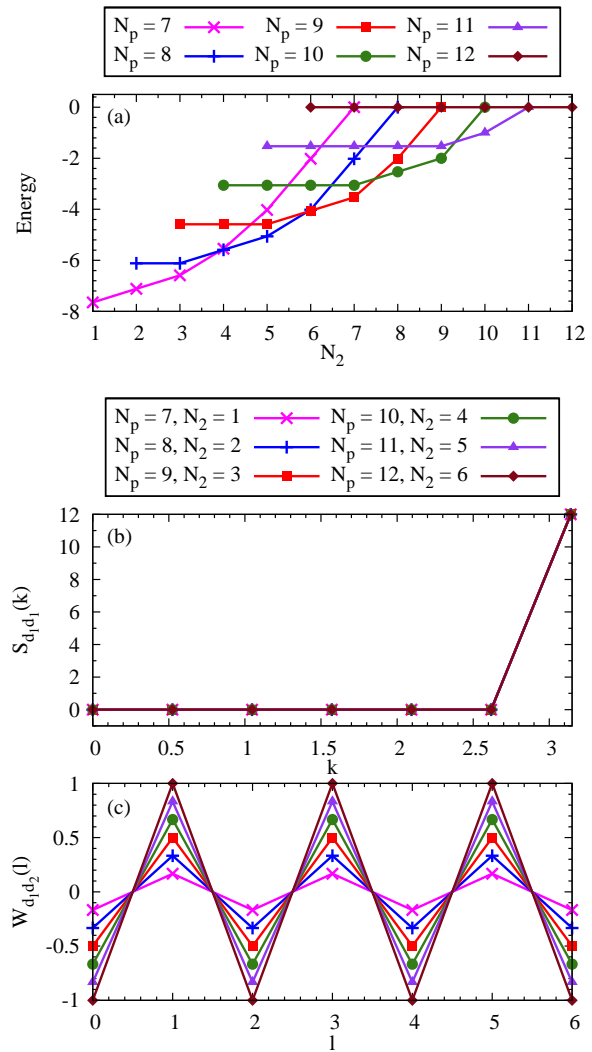}
\caption{(Color online) 
Display  of (a) lowest energy (in units of $\frac{t^2}{2E_p+2V_p} \Big)$
 excluding the polaronic energy;
 (b)  structure factor ${  S_{d_{1}d_{1}}(k)}$; and
 (c) correlation function ${ W_{d_{1}d_{2}}(l)}$
 at filling factor $\nu > 1/2$
in a straight chain with periodic boundary condition.
 The chain has $N=12$ sites and ${ N_p}$ electrons of which ${ N_2}$ are ${ d_2}$-electrons;
 the $d_2$-electrons occupy Jahn-Teller compatible sites just as $d_{z^2}$ electrons in CE-phase
 of PCMO \cite{tokura1}.
The coupling $g=2.3$ and  adiabaticity $\frac{t}{\omega_0}=3.0$.
The structure factor is non-zero only at the reported 
wavevector $k=\pi$ \cite{Milward,braden} with a value $ S_{d_{1}d_{1}}(k=\pi)=4N_1^2/N$;
the correlation function takes the fixed values 
$W_{d_{1}d_{2}}(l_{\rm odd}) = - W_{d_{1}d_{2}}(l_{\rm even}) = 4 N_1N_2/N^2$.
}
\label{fig_nu>0.5}
\end{figure}

{For the situation where the CJT system is above half-filling,}
we display the lowest energies  as a function of 
number of $d_2$-electrons in Fig.~\ref{fig_nu>0.5}(a). 
In the ground state, we observe that the 
electrons occupy both $d_{z^2}$ and $d_{x^2-y^2}$ orbitals
and that the system has degenerate states identifiable by the number ($N_2$) of $d_2$-electrons
in the degenerate state.
For $1/2 < \nu < 1$, the number of degenerate states
in a ground state is given by $N_p-\frac{N}{2}$.
Now, to understand the site occupancy in the ground state,
we note that there is no repulsion between
a pair of either $d_1$-electron and $d_2$-electron or two $d_2$-electrons
[as can be seen in  Eq.~\eqref{eq_eff_ham}].
For the above half-filling case,
in order to avoid the large NN repulsion between $d_1$-electrons,
$\frac{N}{2}$ electrons occupy $d_{z^2}$ orbitals
in a sub-lattice and excess electrons occupy $d_{x^2-y^2}$ orbitals at random sites in the remaining sub-lattice. 
Degenerate states with different values of $N_2$ arise from the fact that a $d_1$-electron sandwiched
between two $d_2$-electrons can be replaced by a $d_2$-electron without altering the energy
when the onsite inter-orbital repulsion is infinite.
This degeneracy keeps the charge
ordering intact, though, it destroys the orbital ordering. 
In the realistic situation of large but finite 
{onsite inter-orbital repulsion}, 
the configuration with the lowest number of $d_2$-electrons yields the lowest energy
as it permits virtual hopping of the $d_1$-electron  to a site with a $d_2$-electron and returning back.

Next, at $\nu > 1/2$ when both $d_1$-electrons and $d_2$-electrons are present,
we study correlation functions and structure factors. If all the $d_1$-electrons
belong to one sub-lattice and the $d_2$ electrons belong to the other sub-lattice, then for all odd values
of $l=l_{\rm odd}$ we obtain $\langle n_{a,j} n_{a,j+l} \rangle = 0$
and for all even 
values of $l=l_{\rm even}$ we get $\langle n_{a,j} n_{b,j+l} \rangle = 0$ with $a \neq b$.
Consequently, Eq.~\eqref{eq_Wuu}
{yields}  
\begin{IEEEeqnarray*}{rCl}
W_{d_{a}d_{a}}(l_{\rm odd}) = - \frac{4 N_a^2}{N^2},
\label{eq_wl_same_red}
\IEEEyesnumber
\end{IEEEeqnarray*}
where $a=1,2$ and 
\begin{IEEEeqnarray*}{rCl}
W_{d_{a}d_{b}}(l_{\rm even}) = - \frac{4 N_a N_b}{N^2},
\label{eq_wl_diff_red}
\IEEEyesnumber
\end{IEEEeqnarray*}
with $a \neq b$.
If the  $d_2$-electrons occur randomly in one sub-lattice,  for
$l=l_{\rm odd}$ and $a \neq b$, we get the simplification 
$\langle n_{a,j} n_{b,j+l} \rangle = 2 \langle n_{a,j}\rangle \langle n_{b,j+l} \rangle$;
then, Eq.~\eqref{eq_Wuu} 
{reduces to}
\begin{IEEEeqnarray*}{rCl}
W_{d_{a}d_{b}}(l_{\rm odd}) =  \frac{4 N_a N_b}{N^2} .
\label{eq_wl_diff_red2}
\IEEEyesnumber
\end{IEEEeqnarray*}
Furthermore, at wavevector $k=\pi$, the structure factor expressed in Eq.~\eqref{eq_Suu} 
simplifies to its  maximum possible value \cite{au_rpys}: 
\begin{IEEEeqnarray*}{rCl}
[S_{d_{a}d_{a}}(k=\pi)] =  \frac{4 N_a^2}{N}.
\label{eq_sk_same_red}
\IEEEyesnumber
\end{IEEEeqnarray*}

We display the structure factor
for $d_1$-electrons in Fig.~\ref{fig_nu>0.5}(b).
The structure factor peaks at the  wavevector $k=\pi$ 
and zero everywhere else indicating a CDW at the reported ordering wavevector $k=\pi$ \cite{Milward,braden};
the  peak value 
{$[S_{d_{1}d_{1}}(k=\pi)=
12$} is in agreement with the
value given by Eq.~\eqref{eq_sk_same_red}.
Hence, above 
half filling, all the $d_1$-electrons reside only in one sub-lattice to avoid large NN repulsion
between them.
Finally, we depict the correlation  between $d_1$-electrons and $d_2$-electrons in Fig.~\ref{fig_nu>0.5}(c).
The function $W_{d_{1}d_{2}}(l)$  oscillates with peaks at odd values of $l$
agreeing with Eq.~\eqref{eq_wl_diff_red2}
and 
{lowest points} at even values of $l$
concurring with  Eq.~\eqref{eq_wl_diff_red}. 
Hence, above half filling, while $d_1$-electrons occupy one sub-lattice,  $d_2$-electrons reside
randomly in the other sub-lattice. Furthermore, since 
{a  NN pair of $d_1$-electron and $d_2$-electron} do not interact [as shown in Eq.~\eqref{eq_eff_ham}],
we note that
 the $d_2$-electrons occupy Jahn-Teller compatible sites just as $d_{z^2}$ electrons in CE-phase 
 of PCMO 
 {at $0.3 < x < 0.5$} \cite{tokura1}.

Therefore, above half filling, system always remains in a CDW state 
and an orbital-density-wave state both with the same ordering wavevector $k=\pi$.

\section{conclusion}
{Transition-metal oxides offer considerable scientific and technological opportunities \cite{millis2}.
An effective Hamiltonian (such as ours) for the CJT effect is a needed building
block for modeling oxides and for aiding material synthesis.}

Duality transformation is a valuable tool in understanding
strongly interacting systems in condensed matter physics, statistical physics, quantum field
theory, and string theory \cite{sachdev,kogut,savit,polchinski}.
{In this work, we demonstrate that the polaronic (Lang-Firsov) transformation 
is actually a duality transformation which  maps
a strong-coupling, many-body problem (where the perturbation 
is proportional to $g\omega_0/t$) to a weak-coupling, tractable many-body problem
(where the small parameter is proportional to $t/(g \omega_0)$). Using perturbation
theory (up to sixth order), we obtain 
{our effective Hamiltonian containing the dominant terms for interactions
at various distances}.}

{Employing our effective  Hamiltonian,}
we find that cooperative Jahn-Teller interaction in two-band manganites R$_{1-x}$A$_x$MnO$_3$
breaks the particle-hole symmetry, i.e., ordering wavevector $k = 2\pi (1-x)$ [$k=\pi$]
for doping fraction 
$x \ge 0.5$ [$x < 0.5$].
Our cooperative picture 
favors a 
{Wigner-crystal order} over bi-stripe order at special fillings $1/3$ and $1/4$, 
thereby shedding light on an existing controversy.
Additionally, at $\nu =1/2$, $1/3$, and $1/4$,
we show that zigzag  chains 
{(pertaining to Wigner-crystal order)}
are  energetically favorable compared to
straight chains; on the other hand, at a lower filling 
{$\nu = 1/5$}, we  demonstrate that 
straight chains pertaining to C-AFM order can be realized. 
{Lastly, even within a strong-coupling picture,
we show for fillings $\nu \neq 1/n$ that electron diffraction patterns
can have 
{finite-peak} intensities at wavevector $k=2\pi\nu$.}

{In future, we would like to apply our approach 
to other transition-metal oxides such as  nickelates, cobaltates, etc.
and study charge stripes at various fillings.}

\section{Acknowledgments}
One of us (S.Y.) acknowledges the hospitality of the TCM group in the Cavendish Laboratory 
(Univ. of Cambridge)
during the initial part of this work.
We thank  P. Littlewood, N. D. Mathur, D. E. Khmelnitskii,
{K. Pradhan,}
and A. Ghosh for useful discussions.

\appendix

\section{General Hamiltonian and Derivation of CJT Model}
\label{app:gen_ham}
The general Hamiltonian for the CJT system in manganites 
can be written as 
$H^G=H_t+H_{ep}+H_l$,
where $H_t$ is the hopping term, $H_{ep}$ the electron-phonon-interaction term,
and $H_l$ the lattice term.
We start with an over-complete basis
$\psi_x = 3x^2-r^2, \psi_y = 3y^2-r^2, \psi_z = 3z^2-r^2$ which satisfies the relation
$\psi_x+\psi_y+\psi_z=0$.
 The basis state $\psi_z$ corresponds to the $d_{z^2}$ orbital
depicted in Fig.\ref{fig:cartoon}.
The hopping term can be expressed in the above basis as:
\begin{IEEEeqnarray}{rCl}
H_t  = &-&t\sum_{i,j,k}[\{d^\dagger_{x^2;i+1,j,k}d_{x^2;i,j,k}+d^\dagger_{y^2;i,j+1,k}d_{y^2;i,j,k} \nonumber \\
& + & d^\dagger_{z^2;i,j,k+1}d_{z^2;i,j,k}\} + {\rm H.c.}], 
\label{eq:hopover}
\end{IEEEeqnarray}
where $ d^\dagger_{x^2;i,j,k},d^\dagger_{y^2;i,j,k}, d^\dagger_{z^2;i,j,k} $ 
 are creation 
 operators
at the site $(i,j,k)$  for $d_{x^2}$, $d_{y^2}$, and $d_{z^2}$ orbitals, respectively. The labeling indices $i$, $j$, and $k$
run along the $x$-, $y$-, and $z$-axes, respectively.
The electron-phonon interaction term can be written as:
\begin{IEEEeqnarray}{rCl}
H_{ep}  = & -& g \omega_0 \sqrt{2 M \omega_0} \sum_{i,j,k} [n_{x^2;i,j,k} Q_{x;i,j,k} \nonumber \\
& + & n_{y^2;i,j,k} Q_{y;i,j,k} + n_{z^2;i,j,k} Q_{z;i,j,k} ],
\label{eq:elphover}
\end{IEEEeqnarray}
where $g$ is the electron-phonon coupling, $M$ is the mass of an oxygen ion,  $\omega_0$ is the frequency of optical phonons, 
and $n_{x^2(y^2,z^2);i,j,k} = d^\dagger_{x^2(y^2,z^2);i,j,k} d_{x^2(y^2,z^2);i,j,k} $ are the number operators.
Furthermore, $Q_{x;i,j,k}$, $Q_{y;i,j,k}$ and $Q_{z;i,j,k}$ are defined in terms of the displacements
[$u_{x;i,j,k}~ \& ~u_{x;i-1,j,k}$; $ u_{y;i,j,k}~ \&~ u_{y;i,j-1,k}$; $ u_{z;i,j,k} ~\&~ u_{z;i,j,k-1}$]
of oxygen ions 
around (and in the direction of) the $d_{x^2}$, $d_{y^2}$, and $d_{z^2} $ orbitals, respectively, as follows:
$Q_{x;i,j,k} = u_{x;i,j,k}-u_{x;i-1,j,k}$, $Q_{y;i,j,k} = u_{y;i,j,k}-u_{y;i,j-1,k}$, and $Q_{z;i,j,k} = u_{z;i,j,k}-u_{z;i,j,k-1}$.
Here, besides considering the displacement of the ions, we also consider their kinetic energy, thereby 
invoking quantum nature of the phonons. Then,
the lattice Hamiltonian is given by
\begin{IEEEeqnarray}{rCl}
H_l  = && \frac{M}{2} \sum_{i,j,k} [ \dot{u}^2_{x;i,j,k} + \dot{u}^2_{y;i,j,k} + \dot{u}^2_{z;i,j,k} ] \nonumber \\ 
& + & \frac{K}{2} \sum_{i,j,k} [ u^2_{x;i,j,k} + u^2_{y;i,j,k} + u^2_{z;i,j,k} ], 
\label{eq:gen_lattice}
\end{IEEEeqnarray}
where $ \dot{u}_{x;i,j,k} $, $ \dot{u}_{y;i,j,k} $, and $ \dot{u}_{z;i,j,k} $ are the time derivatives
 of the oxygen-ion displacements 
$u_{x;i,j,k}$, $ u_{y;i,j,k}$, and $ u_{z;i,j,k}$, respectively.

 The usual orthogonal basis states $\psi_{x^2-y^2}$ and $\psi_{z^2}$ are related to the over-complete basis states
$\psi_x$, $\psi_y$, and $\psi_z$ as follows:
\begin{IEEEeqnarray*}{rCl}
&&\psi_{x^2-y^2} = \frac{1}{\sqrt3} (\psi_x-\psi_y), \\
&&\psi_{z^2} = \psi_z.
\label{eq:psi_x}
\\*\IEEEyesnumber
\end{IEEEeqnarray*}
From Eq.~\eqref{eq:psi_x} we get,
\begin{IEEEeqnarray}{rCl}
\psi_x &=& \frac{1}{2} (\sqrt{3} \psi_{x^2-y^2}-\psi_{z^2}), \nonumber \\
\psi_y &=& - \frac{1}{2} (\sqrt{3} \psi_{x^2-y^2}+\psi_{z^2}), \nonumber \\
\psi_z &=& \psi_{z^2}. \nonumber\\
\label{eq:psiz^2}
\end{IEEEeqnarray}

 Next, using Eq.~\eqref{eq:psiz^2}, we express the general Hamiltonian in  the  orthogonal basis $\psi_{x^2-y^2}$ and $\psi_{z^2}$ as follows:
\begin{widetext}
\begin{IEEEeqnarray}{rCl}
\label{eq:gen_hop_com}
H_t = 
&-&\frac{t}{4} \sum_{i,j,k} \{(d^\dagger_{z^2;i+1,j,k},d^\dagger_{x^2-y^2;i+1,j,k}) 
\begin{pmatrix}
1 & -\sqrt{3} \\
-\sqrt{3} & 3
\end{pmatrix}
\begin{pmatrix}
d_{z^2;i,j,k} \\
d_{x^2-y^2;i,j,k}        
\end{pmatrix} 
+ {\rm H. c.} \}  
-\frac{t}{4} \sum_{i,j,k} \{ (d^\dagger_{z^2;i,j+1,k},d^\dagger_{x^2-y^2;i,j+1,k}) \nonumber \\
&\times&
\begin{pmatrix}
1 & \sqrt{3} \\
\sqrt{3} & 3
\end{pmatrix}
\begin{pmatrix}
d_{z^2;i,j,k} \\
d_{x^2-y^2;i,j,k}        
\end{pmatrix} 
+ {\rm H. c.} \} 
-t\sum_{i,j,k} \{(d^\dagger_{z^2;i,j,k+1},d^\dagger_{x^2-y^2;i,j,k+1}) 
\begin{pmatrix}
1 & 0 \\
0 & 0
\end{pmatrix} 
\begin{pmatrix}
d_{z^2;i,j,k} \\
d_{x^2-y^2;i,j,k}        
\end{pmatrix}
+ {\rm H. c.} \},
\end{IEEEeqnarray}
\begin{IEEEeqnarray}{rCl}
H_{ep} = &-&\frac{1}{4} g \omega_0 \sqrt{2 M \omega_0} \nonumber \\
&\times&\sum_{i,j,k} (d^\dagger_{z^2;i,j,k},d^\dagger_{x^2-y^2;i,j,k}) 
\begin{pmatrix}
Q_{x;i,j,k}+Q_{y;i,j,k}+4Q_{z;i,j,k} & -\sqrt{3}Q_{x;i,j,k}+\sqrt{3}Q_{y;i,j,k} \\
-\sqrt{3}Q_{x;i,j,k}+\sqrt{3}Q_{y;i,j,k} & 3Q_{x;i,j,k}+3Q_{y;i,j,k}
\end{pmatrix}
\begin{pmatrix}
d_{z^2;i,j,k} \\
d_{x^2-y^2;i,j,k}        
\end{pmatrix} ,
\label{eq:gen_elph_com}
\end{IEEEeqnarray}
and $H_l$ is again given by Eq. (\ref{eq:gen_lattice}).
Here, it should be mentioned that an expression for $H^G$ in an alternate basis has been derived  in Ref. \onlinecite{au:allen}; 
however, these authors consider classical phonons. 
\end{widetext}

Now, we consider a one-dimensional  Jahn-Teller chain with cooperative electron-phonon 
interaction along the z-direction and non-cooperative  electron-phonon interaction (of the Holstein-type \cite{au_holstein,au_sdadys})
along the x- and
y-directions as shown in Fig.~\ref{fig:cartoon} of the main text. The lattice term given by Eq.~\eqref{eq:gen_lattice} can be 
written for this case as follows:
\begin{IEEEeqnarray*}{rCl}
H^{CJT}_l  = &&\> \frac{M}{2} \sum_{k} [\dot{u}^2_{x;0,k}+\dot{u}^2_{x;1,k} + \dot{u}^2_{y;0,k}+\dot{u}^2_{y;1,k}\\
&&\>+ \dot{u}^2_{z;k}]
+\frac{K}{2} \sum_{k} [ u^2_{x;0,k}+u^2_{x;1,k} + u^2_{y;0,k}\\
&&\>+u^2_{y;1,k} + u^2_{z;k} ].\IEEEyesnumber 
\label{eq:1dlatticeintermediate}
\end{IEEEeqnarray*}
We define $Q^{\prime}_{x;k} \equiv u_{x;1,k}+u_{x;0,k}$, $Q^{\prime}_{y;k} \equiv u_{y;1,k}+u_{y;0,k}$,
$Q_{x;k} \equiv u_{x;1,k}-u_{x;0,k}$, and $Q_{y;k} \equiv u_{y;1,k}-u_{y;0,k}$ 
and incorporate these definitions in 
Eq.~\eqref{eq:1dlatticeintermediate} to obtain
\begin{IEEEeqnarray*}{rCl}
&& \!\! H^{CJT}_l \\
&& ~ =  \frac{M}{2} \sum_{k} \bigg [ \frac{1}{2}\{\dot{Q}^{\prime 2}_{x;k}+\dot{Q}^{\prime 2}_{y;k}\}
+\frac{1}{4}\{\dot{Q}^{+2}_{xy;k} 
 + \dot{Q}^{-2}_{xy;k}\}+ \dot{u}^2_{z;k} \bigg ] \\
&& ~~ +\frac{K}{2} \sum_{k} \bigg [\frac{1}{2} \{Q^{\prime 2}_{x;k}+Q^{\prime 2}_{y;k}\} 
 + \frac{1}{4}\{Q^{+2}_{xy;k}+Q^{-2}_{xy;k}\}+u^2_{z;k} \bigg ], \\
\IEEEyesnumber 
\label{eq:1dlattice}
\end{IEEEeqnarray*}
where $Q^\pm_{xy;k} \equiv Q_{x;k} \pm Q_{y;k} $.
For the  present single-chain case,  Eqs.~\eqref{eq:gen_hop_com} and
\eqref{eq:gen_elph_com} reduce to the following equations:
\begin{IEEEeqnarray*}{rCl}
H^{CJT}_t = &-&t \sum_k (d^\dagger_{z^2;k+1}d_{z^2;k}+ {\rm H.c.}) ,
\IEEEyesnumber
\label{eq:HCJT_t}
\end{IEEEeqnarray*}
and 
\begin{IEEEeqnarray*}{rCl}
&&\!\!\!\! \frac{H^{CJT}_{ep}}{g \omega_0 \sqrt{2 M \omega_0} } \\
&&  = -\sum_k \bigg [ \Big \{(u_{z;k}-u_{z;k-1})
+\frac{1}{4}Q^+_{xy;k} \Big \} d^\dagger_{z^2;k}d_{z^2;k} \\
&& \qquad \qquad + \frac{3}{4}Q^+_{xy;k} d^\dagger_{x^2-y^2;k}d_{x^2-y^2;k} \\
&& \qquad \qquad -\frac{\sqrt{3}}{4} Q^-_{xy;k} \Big ( d^\dagger_{z^2;k}d_{x^2-y^2;k} + {\rm H.c.} \Big ) \bigg ] . 
\IEEEyesnumber
\label{eq:HCJT_ep}
\end{IEEEeqnarray*}
Next, we note that the center-of-mass displacement terms $Q^{\prime }_{x;k}$ and $Q^{\prime }_{y;k}$ as well as
the center-of-mass momentum terms $\dot{Q}^{\prime }_{x;k}$ and $\dot{Q}^{\prime }_{y;k}$ 
 of 
 { Eq. \eqref{eq:1dlattice}}
do not couple to the electrons 
 [as can be seen from Eqs. (\ref{eq:HCJT_t}) and \eqref{eq:HCJT_ep}]. Hence, for our single-chain case, 
Eq. (\ref{eq:1dlattice}) simplifies to be
{
\begin{IEEEeqnarray*}{rCl}
H^{CJT}_l &=& 
\sum_k\left[\frac{1}{2} M  {{\dot{u}}_{z;k}}^2 + \frac{1}{2} K  u^2_{z;k}\right] \\
&& +\sum_k
\left[
\frac{1}{2}
\frac{M}{4} {{\dot{Q}^{+2}}_{xy;k}} + 
\frac{1}{2}
\frac{K}{4} Q^{+2}_{xy;k}\right]  \\
&& + \sum_k
\left[
\frac{1}{2}
\frac{M}{4}  {\dot{Q}^{-2}_{xy;k}} + 
\frac{1}{2}
\frac{K}{4} Q^{-2}_{xy;k}\right] .
\IEEEyesnumber\label{eq:HCJT_l}
\end{IEEEeqnarray*}
}
The general Hamiltonian for the present CJT single chain  can be expressed as follows by adding Eqs. \eqref{eq:HCJT_t}, \eqref{eq:HCJT_ep}, and
\eqref{eq:HCJT_l}:
\begin{IEEEeqnarray}{rCl}
 H^{CJT}=H^{CJT}_t+H^{CJT}_{ep}+H^{CJT}_{l} .
\label{eq:HCJT}
\end{IEEEeqnarray}
 Next, by using the following second-quantized representation of the various displacement operators:
\begin{IEEEeqnarray*}{rCl}
u_{z;k} = \frac{a^\dagger_{z;k}+a_{z;k}}{\sqrt{2M\omega_0}},
Q^+_{xy;k}= \frac{b^\dagger_k+b_k}{\sqrt{2\frac{M}{4}\omega_0}},
Q^-_{xy;k}= \frac{c^\dagger_k+c_k}{\sqrt{2\frac{M}{4}\omega_0}},
\end{IEEEeqnarray*}
in the above Hamiltonian of Eq. (\ref{eq:HCJT}), we obtain
\begin{IEEEeqnarray}{rCl}
\!\!\!\!\!\!\!\! H^{CJT} =& -&t \sum_k (d^\dagger_{z^2;k+1}d_{z^2;k}+ {\rm H.c.}) \nonumber \\ 
&-&g \omega_0 \sum_k \bigg [(a^\dagger_{z;k}+a_{z;k})(n_{z^2;k}-n_{z^2;k+1}) \nonumber \\
&&\qquad ~~~~+  \frac{1}{2} (b^\dagger_k+b_k) (n_{z^2;k}+3 n_{x^2-y^2;k} ) \nonumber \\
&& \qquad ~~~~ - \frac{\sqrt{3}}{2} (c^\dagger_k+c_k) (d^\dagger_{z^2;k}d_{x^2-y^2;k}+ {\rm H.c.})\bigg ] \nonumber \\
&+& \omega_0 \sum_k (a^\dagger_{z;k}a_{z;k} + b^\dagger_kb_k + c^\dagger_kc_k), 
\label{appeq:HCJT_sq}
\end{IEEEeqnarray}
where $n_{z^2;k} \equiv d^\dagger_{z^2;k}d_{z^2;k} $ and $n_{x^2-y^2;k} \equiv d^\dagger_{x^2-y^2;k}d_{x^2-y^2;k} $.

\section{Perturbation up to second-order}
\label{app:IIorder}
We  adopt a polaronic transformation for the Hamiltonian in Eq. (\ref{eq_HCJT_sq}) of the main
text so that we can
perform perturbation theory. In the polaronic frame of reference, the transformed Hamiltonian  
 reads $\tilde{H}^{CJT} = \exp(S) H^{CJT} \exp(-S)$
{where the operator $S$ is defined in Eq. (\ref{op_lang}) of the main text.}
Then, the transformed Hamiltonian can be expressed  as $\tilde{H}^{CJT} = H_0 + H_1$ with
\begin{IEEEeqnarray}{rCl}
H_0 &=& \omega_0 \sum_k (a^\dagger_{k}a_{k} + b^\dagger_k b_k + c^\dagger_k c_k)
- \frac{9}{4} g^2 \omega_0 \sum_k (n_{1,k}  
+ n_{2,k}) \nonumber  \\
&&-\frac{3}{2} g^2 \omega_0 \sum_k n_{1,k} n_{2,k} 
+ 2 g^2 \omega_0 \sum_k n_{1,k}n_{1,k+1} \nonumber \\  
&&- t e^{-\frac{13}{4}g^2} \sum_k (d^\dagger_{1,k+1}d_{1,k}+ {\rm H.c.}),
\label{eq_HA_0}
\end{IEEEeqnarray}
where the term $2 g^2 \omega_0 \sum_k n_{1,k}n_{1,k+1}$ arises because of the cooperative nature of the interaction;
furthermore,
the attractive interaction term $-\frac{3}{2} g^2 \omega_0 \sum_k n_{1,k} n_{2,k}$ will be negated
by a much larger repulsive Coulombic term  $U \sum_k n_{1,k} n_{2,k}$ because of which no site 
can have both the orbitals 
occupied simultaneously. 
{Here it is important to point out that there is no interaction between
a NN pair of either $d_1$-electron and $d_2$-electron or two $d_2$-electrons.
}
The remaining term of $\tilde{H}^{CJT}$ can be written as
 $H_1 \equiv H^I_1+ H^{II}_1$ with 
\begin{IEEEeqnarray}{rCl}
H^I_1 
= -t e^{-\frac{13}{4} g^2}
 \sum_k [ d^\dagger_{1,k+1}d_{1,k}
\{
{\cal{T}}_{+}^{k \dagger} {\cal{T}}^{k}_{-}
-1 \} + {\rm H.c.}],
\IEEEeqnarraynumspace
\label{eq_HI}
\end{IEEEeqnarray}
where ${\cal{T}}^{k}_{\pm} \equiv \exp[\pm g( 2 a_{k} - a_{k-1} - a_{k+1})\pm \frac{g}{2}(b_k-b_{k+1})]$ 
and 
\begin{IEEEeqnarray}{rCl}
H^{II}_1 
&=& \frac{\sqrt{3}}{2} g \omega_0 e^{-\frac{3}{2}g^2} \sum_k (c^\dagger_k+c_k)\bigg [d^\dagger_{1,k}d_{2,k}
 \nonumber \\
&&\times\>e^{g(a^\dagger_{k-1}-a^\dagger_{k}+b^\dagger_k)} e^{-g(a_{k-1}-a_{k}+b_k)} + {\rm H.c.}\bigg].
\IEEEeqnarraynumspace
\label{eq_HII}
\end{IEEEeqnarray}
Now, to perform perturbation theory, we note that the eigenstates of $H_0$ are given 
by $|n,m\rangle =|n\rangle_{el}\otimes |m\rangle_{ph}$
with $|0,0\rangle$ being the ground state. 
We consider the case when
the coefficients of the perturbation terms $H^I_1$ and $H^{II}_1$ in Eqs. (\ref{eq_HI}) and (\ref{eq_HII}), respectively, 
satisfy the conditions $t e^{-\frac{13}{4} g^2} << \omega_0$ 
and 
{$\frac{\sqrt{3}}{2} g  e^{-\frac{3}{2}g^2} << 1$}
Now, the
first order correction is zero and the
second-order perturbation term [obtained using Schrieffer-Wolff transformation
 as mentioned in Eq. (6) of Ref. \onlinecite{au_rpys}]
is expressed as
\begin{IEEEeqnarray}{rCl}
H^{(2)} = \sum_{m}
\frac{\langle 0|_{ph} H_{1} |m\rangle_{ph}
 \langle m|_{ph} H_{1} |0\rangle_{ph}}
{E_{0}^{ph} - E_{m}^{ph}} .
\label{H^2}
 \end{IEEEeqnarray}   
In  Eq.~\eqref{H^2}, the contribution of cross terms involving  $H^I_1$ and $H^{II}_1$ is zero because the phonons 
do not match; hence, we get 
\begin{IEEEeqnarray}{rCl}
H^{(2)} =&& \sum_{m}
\frac{\langle 0|_{ph} H^I_{1} |m\rangle_{ph}
 \langle m|_{ph} H^I_{1} |0\rangle_{ph}}
{E_{0}^{ph} - E_{m}^{ph}} \nonumber \\
&+&\sum_{m} \frac{\langle 0|_{ph} H^{II}_{1} |m\rangle_{ph}
 \langle m|_{ph} H^{II}_{1} |0\rangle_{ph}}
{E_{0}^{ph} - E_{m}^{ph}}.
\label{H^2_1}
 \end{IEEEeqnarray} 
We will first evaluate the term involving $H^{II}_1$ in the above equation. After some algebra, we get the following expression:
 \begin{IEEEeqnarray}{rCl}
&& \!\!\!\! \sum_{m} \frac{\langle 0|_{ph} H^{II}_{1} |m\rangle_{ph}
 \langle m|_{ph} H^{II}_{1} |0\rangle_{ph}}
{E_{0}^{ph} - E_{m}^{ph}} \nonumber \\
&& \approx -\frac{\omega_0}{4} \sum_k \Big [ n_{1,k}  
+ n_{2,k} 
- 2 n_{1,k} n_{2,k} \Big ] .
\label{H^II}
 \end{IEEEeqnarray} 
We note that the coefficients of the  terms $n_{1,k}$, $ n_{2,k}$, and $ n_{1,k} n_{2,k}$
in the above equation are much smaller than the coefficients of 
the same terms in Eq. (\ref{eq_HA_0}); consequently, we ignore the contribution from 
Eq. (\ref{H^II}) in the expression for the effective Hamiltonian of the CJT chain.  

Next, we evaluate the term involving $H^{I}_1$ in Eq.~\eqref{H^2_1} and obtain:
\begin{widetext}
\begin{IEEEeqnarray*}{rCl}
&& \sum_{m}
\frac{\langle 0|_{ph} H^I_{1} |m\rangle_{ph}
 \langle m|_{ph} H^I_{1} |0\rangle_{ph}}
{E_{0}^{ph} - E_{m}^{ph}} \\
&& = -\frac{t^2}{\omega_0}e^{-\frac{13}{2} g^2}
G_3 \left(2,2,\frac{1}{4}\right) 
\sum_k \left [
d^\dagger_{1,k+2}(1-n_{1,k+1})(1-n_{2,k+1})d_{1,k}
+ {\rm H.c.}\right ] \\
&& \quad - \frac{t^2}{\omega_0}e^{-\frac{13}{2} g^2}
 G_5 \left(4,1,1,\frac{1}{4},\frac{1}{4}\right)
\sum_k \big[ n_{1,k} (1-n_{1,k-1}) (1-n_{2,k-1}) (1- n_{1,k-2})
+ n_{1,k} (1-n_{1,k+1}) (1-n_{2,k+1}) (1- n_{1,k+2} ) 
\big] \\
&& \quad - (.) \sum_k \big[ 
n_{1,k} (1-n_{1,k-1}) (1-n_{2,k-1})  n_{1,k-2}
+ n_{1,k} (1-n_{1,k+1}) (1-n_{2,k+1})  n_{1,k+2} 
\big],
\IEEEyesnumber
\label{eq_H^2_1_1}
\end{IEEEeqnarray*}
{where the coefficient [denoted by (.)] of the last term on the RHS  will be given below and}
$G_n(\alpha_1,\alpha_2,\ldots,\alpha_n)
{\approx}
\frac{e^{\sum_{i=1}^n\alpha_ig^2}}{\sum_{i=1}^n\alpha_ig^2} $ 
for large values of $g^2$ (see Appendix \ref{app:serieses} for details).

Then, on using the above approximation for $G_n(\alpha_1,\alpha_2,\ldots,\alpha_n)$
at large $g^2$,  Eq. (\ref{eq_H^2_1_1}) simplifies as follows: 
\begin{IEEEeqnarray*}{rCl}
&& \sum_{m}
\frac{\langle 0|_{ph} H^I_{1} |m\rangle_{ph}
 \langle m|_{ph} H^I_{1} |0\rangle_{ph}}
{E_{0}^{ph} - E_{m}^{ph}} \\
&& = -\frac{4}{17} \frac{t^2}{g^2 \omega_0}e^{-\frac{9}{4} g^2} 
\sum_k \left [
d^\dagger_{1,k+2}(1-n_{1,k+1})(1-n_{2,k+1})d_{1,k}
+ {\rm H.c.}\right ] \\
&& \quad - \frac{2}{13} \frac{t^2}{g^2 \omega_0}
\sum_k
\big[ n_{1,k} (1-n_{1,k-1}) (1-n_{2,k-1}) (1- n_{1,k-2})
+ n_{1,k} (1-n_{1,k+1}) (1-n_{2,k+1}) (1- n_{1,k+2}) 
\big] \\
&& \quad - \bigg( \frac{t^2}{2E_p+4V_p} \bigg ) \sum_k 
\big[ n_{1,k} (1-n_{1,k-1}) (1-n_{2,k-1})  n_{1,k-2}
+ n_{1,k} (1-n_{1,k+1}) (1-n_{2,k+1})  n_{1,k+2} 
 \big] ,
\IEEEyesnumber
\label{eq_HA^2_1_2}
\end{IEEEeqnarray*}
where the coefficients of the first and second terms on the RHS
agree with the corresponding terms in Eq.~\ref{eq_H^2_1_2}
in the main text; then, the coefficient of the  
last term on the RHS is identified from   Eq.~\ref{eq_H^2_1_2} in the main text.
\end{widetext}

\section{Simplification of  the function $G_n(\alpha_1,\alpha_2,\ldots,\alpha_n)$}
\label{app:serieses}
In this appendix, we obtain simple expressions for the function $G_n(\alpha_1,\alpha_2,\ldots,\alpha_n)$
appearing in  Appendix \ref{app:IIorder}.
The general term  $G_n(\alpha_1,\alpha_2,\ldots,\alpha_n)$ 
is defined as 
\begin{IEEEeqnarray*}{rCl}
G_n(\alpha_1,\alpha_2,\ldots,\alpha_n) &\equiv& F_n(\alpha_1,\alpha_2,\ldots,\alpha_n)\\
&&+\sum_{k=1}^{n-1}\sum_c
F_k(\alpha_{c_1},\alpha_{c_2},\ldots,\alpha_{c_k}) , 
 \end{IEEEeqnarray*}   
where 
\begin{IEEEeqnarray*}{rCl}
\!\!\!\! F_n(\alpha_1, \ldots , \alpha_n ) \equiv \sum_{m_1=1}^{\infty} 
 ...
\sum_{m_n=1}^{\infty}
 \frac {(\alpha_1 g^2)^{m_1} \ldots (\alpha_n g^2)^{m_n}}
{m_1!\ldots m_n!(m_1+ \ldots + m_n)} ,
 \end{IEEEeqnarray*}   
and the summation over $c$ represents summing over all possible
$^nC_k$ combinations of $k$ arguments chosen from the total set of 
 $n$ arguments $\{\alpha_1,\alpha_2,\ldots,\alpha_n\}$.

We evaluate the derivative of the general term
$G_n(\alpha_1,\alpha_2,\ldots,\alpha_n)$
with respect to  $g^2$:
\begin{IEEEeqnarray*}{rCl}
&&g^2\frac{d}{dg^2}G_n(\alpha_1,\alpha_2,\ldots,\alpha_n)
\\
&&\quad=(e^{\alpha_1g^2}-1)(e^{\alpha_2g^2}-1)\ldots(e^{\alpha_ng^2}-1)\\
&&\qquad +\sum_{k=1}^{n-1}\sum_c (e^{\alpha_{c_1}g^2}-1)(e^{\alpha_{c_2}g^2}-1)\ldots(e^{\alpha_{c_k}g^2}-1)\\
&&\quad=\left [ \Pi_{i=1}^{n}\big \{(e^{\alpha_ig^2}-1)+1\big \}\right ]-1\\
&&\quad =e^{\sum_{i=1}^n\alpha_ig^2}-1. \IEEEyesnumber
\end{IEEEeqnarray*}
Then, the general term is obtained to be 
\begin{IEEEeqnarray*}{rCl}
G_n(\alpha_1,\alpha_2,\ldots,\alpha_n)
&=&\int \frac{e^{\sum_{i=1}^n\alpha_ig^2}-1}{g^2}dg^2\\
&=& \int \sum^\infty_{m=1} \frac{(\sum_{i=1}^n\alpha_i)^m(g^2)^{(m-1)}}{m!}dg^2\\
&=& \sum^\infty_{m=1} \frac{(\sum_{i=1}^n\alpha_ig^2)^m}{m\>m!}.\IEEEyesnumber
\end{IEEEeqnarray*}
For large values of $g^2$, we have the approximation
\begin{IEEEeqnarray*}{rCl}
\int \frac{e^{\sum_{i=1}^n\alpha_ig^2}-1}{g^2}dg^2
\approx\frac{e^{\sum_{i=1}^n\alpha_ig^2}}{\sum_{i=1}^n\alpha_ig^2}.
\end{IEEEeqnarray*}

\section{Depiction of fourth-order processes}
\label{IV-process}

\begin{figure}
\vspace{0.8cm}
\includegraphics[scale=1,angle=0]{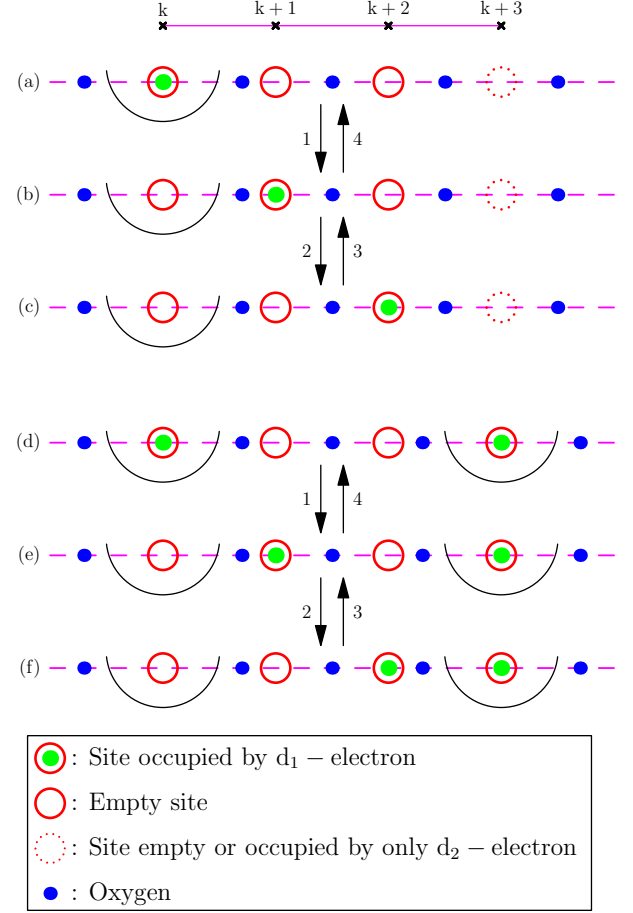} 
\caption{(Color online) Display of initial/final and intermediate states with concomitant
lattice distortions in a fourth-order perturbation
process involving electron hopping  right to NNN site and returning. 
When NNNN site is unoccupied by $d_1$-electron, we depict 
 initial/final state (a);  intermediate states (b) and (c).
When NNNN site is occupied by $d_1$-electron, we show 
 initial/final state (d);  intermediate states (e) and (f).
The numbered
arrows indicate the order of hopping.
}
\label{fig_fourthorder}
\end{figure}

\pagebreak

\end{document}